\documentclass{aastex}
\input{epsf.sty}
\usepackage{emulateapj5}

\def\kms{km\thinspace s$^{-1}$}     
\def\deg{\ifmmode^\circ\else$^\circ$\fi}  
\def\arcs{\ifmmode {'' }\else $'' $\fi}  
\def\arcm{\ifmmode {' }\else $' $\fi}    

\def\msolar{M$_\odot$}
\def\lya{Ly-$\alpha$}
\def\cm2{cm$^{-2}$}
\def\NHI{N$_{HI}$}
\def\lNHI{log N$_{HI}$\ }
\newcommand{\gapprox}{\ifmmode \buildrel > \over {_\sim} \else $\buildrel >\over {_\sim}$\fi}
\newcommand{\lapprox}{\ifmmode \buildrel < \over {_\sim} \else $\buildrel <\over {_\sim}$\fi}
\def\h70{$h^{-1}_{70}$}

\begin{document}

\title{Probing the Size of Low-Redshift \lya\ Absorbers} 

\author{Rosenberg, Jessica L.}
\affil{Center for Astrophysics \& Space Astronomy, Department of Astrophysical
and Planetary Sciences, University of Colorado, Boulder, CO 80309}
\author{Ganguly, Rajib} 
\affil{Space Telescope Science Institute, 3700 San Martin Drive, 
Baltimore, MD 21218}
\author{Giroux, Mark L.}
\affil{Dept. of Physics and Astronomy, East Tennessee State University, Johnson
City, TN 37614}
\author{Stocke, John T.}
\affil{Center for Astrophysics \& Space Astronomy, Department of Astrophysical
and Planetary Sciences, University of Colorado, Boulder, CO 80309}

\begin{abstract}
The 3C~273 and RX J1230.8+0115 sight lines probe the outskirts of the Virgo 
Cluster at physical separations between the sight lines of 200-500\h70\ kpc. We 
present an analysis of 
available HST STIS echelle and FUSE UV spectroscopy of RX J1230.8+0115, in which 
we detect five \lya\ absorbers at Virgo distances. One of these absorbers is a
blend of two strong metal line absorbers at a recession velocity coincident with 
the highest neutral hydrogen column density absorber in the 3C~273 sight line, 
$\sim$ 350 \h70\ 
kpc away. The consistency of the metal line column density ratios in the RX 
J1230.8+0115 sight line allows us to determine the ionization
mechanism (photoionization) for these absorbers. While the low 
signal-to-noise ratio of the FUSE spectrum limits our ability to model the 
neutral hydrogen column density of these absorbers precisely, we are able to 
constrain them to be in the range 10$^{16-17}$ \cm2. The properties of these 
absorbers are similar to those obtained for the nearby 3C~273 absorber studied 
by Tripp and collaborators. However, the inferred line-of-sight size for the 
3C~273 absorber is only 70 pc, much smaller than those inferred in 
RX J1230.8+0115, which are 10 -- 30 \h70\ of kpc. The small sizes of all three 
absorbers are at odds 
with the \gapprox 350 \h70\ kpc minimum transverse size implied by an 
application of the standard QSO line pairs analysis. On the basis of absorber 
associations between these two sight lines 
we conclude that a large-scale structure filament produces a correlated, not 
contiguous, gaseous structure in this region of the Virgo Supercluster. These
data may indicate that we are detecting overdensities in the large scale 
structure filaments in this region. Alternatively, the presence of a galaxy 
71\h70\ kpc from a 3C 273 absorber may indicate that we have probed outflowing, 
starburst driven shells of gas associated with nearby galaxies.

\end{abstract}

\section{Introduction}

The structure and origin of \lya\ ``forest'' absorbers remains uncertain despite 
decades of study with ground-based telescopes and more recently with the Hubble 
Space Telescope (HST). It has been argued that \lya\ forest absorbers are related 
to the extended halos of bright galaxies \citep[e.g.,][]{lin:00,lanzetta:95}, 
that they are associated with dwarf or low surface brightness galaxies 
\citep[e.g.,][]{impey:97}, or that they are one phase of the gaseous medium in 
spiral-rich galaxy groups \citep{morris:94,shull:98,tripp:01,bowen:02}. 
Alternatively, it has been proposed that they are ``primordial'' overdensities 
associated with large scale structure, but not with any galaxy in particular 
\citep{dave:99,stocke:01,penton:02}. It is possible that all of these 
scenarios are represented. Therefore, knowing the fraction of the \lya\ forest
associated with a given scenario, as a function of column 
density and redshift, is essential if we are going to use studies of the \lya\
forest to understand the formation and early evolution of galaxies. The size and
shape of these absorbing structures bears directly on the fraction of absorbers
associated with a specific environment.

The correlation between absorbers in quasar pairs is often used to determine 
statistically the sizes of absorbers, particularly at high redshift
\citep[e.g.,][]{Foltz:84,Bechtold:94,crotts:94,Dinshaw:97,Dinshaw:98}. These 
QSO sight line pair experiments yield characteristic radii for  
higher column density (\lNHI $\geq$ 14.5 \cm2) absorbers of 
300 -- 700\h70\ kpc, with some evidence for increasing sizes at 
lower redshifts \citep{Dinshaw:98}. These very large size measurements are 
consistent with the expectations of cosmological N-body simulations 
\citep{dave:99} as well as the expectations from an analytical Jeans' mass theory 
of gravitationally-bound, photoionized clouds \citep{schaye:01}. From a census
of \lya\ forest absorbers using HST/GHRS and STIS observations, 
\citet{penton:00b,penton:03} find that $\sim$40\% of all local baryons reside in
the forest. This estimate, however, assumes a characteristic radii for spherical
absorbing structures of 100\h70\ kpc. It is uncertain whether this size, derived 
from high redshift absorber pairs, is an appropriate measurement of the absorber's
contiguous size or merely the length scale over which the absorbers are
associated with the same large scale structure (Dinshaw et al. 1998).

Alternatively, the physical line-of-sight dimension of absorbers can be
derived by assuming photoionization equilibrium, from measurements of H~I column
density, metal line strengths and the ratio of line strengths for different 
ionization states of the same species. The line strengths and ratios can be used 
to constrain metallicities and physical densities in these structures
assuming photoionization equilibrium with the extragalactic ionizing flux 
\citep{bechtold:94, shull:99, Weymann:02, Donahue:95}. The calculation of 
hydrogen densities from the photoionization models
can then be used to infer physical sizes along the line-of-sight. The sizes
determined for absorbers with strong, low ionization metal lines like C~II, Si~II, 
Mg~II and Fe~II \citep{Steidel:95, rigby:02, tripp:02}, are much smaller than the 
characteristic sizes in the plane of the sky obtained from the QSO pairs 
experiments. For example, the low ionization absorbers with relatively weak 
Mg~II absorption ($<$0.3\AA) have sizes derived from a photoionization 
equilibrium calculation which are $<$ 1 kpc \citep{rigby:02}. 

In this paper we present the case of a pair of QSOs, 3C~273 and RX J1230.8+0115,  
which probe the outskirts of the Virgo Cluster at physical separations between
the sight lines of 200 -- 500\h70\ kpc. Both sight lines contain strong (\lNHI 
$\sim$ 16 \cm2) 
\lya\ absorption systems which have relatively low ionization absorption line 
spectra. This low redshift pair provides us the opportunity to use both of the
aforementioned size determination methods at redshifts where we can obtain a
deep galaxy census.
 
Tripp et al. (2002; T2002 hereafter) presented an analysis of the absorption 
along the line-of-sight to 3C~273, 0.9\deg\ away from RX J1230.8+0115 on the sky. 
T2002 studied the two strong absorbers at Virgo
distances ($cz$=1011 and 1586 \kms) in a STIS echelle (7 \kms\
resolution) spectrum of 3C~273 obtained by \citet{Heap:02}. These two absorbers 
had been detected and studied using the Faint Object Spectrograph \citep{Bahcall:91} 
and the GHRS \citep{Morris:91,Weymann:95}. More recently 3C~273 was observed 
with the Far-Ultraviolet Spectroscopic Explorer \citep[FUSE;][]{Sembach:01} 
allowing detections of the higher Lyman lines in both of these Virgo absorbers 
and O~VI absorption in the 1011 \kms\ absorber. The 1586 \kms\ absorber contains 
only low ionization absorption lines of 
C~II, Si~II and Si~III. Using a curve-of-growth analysis and a standard 
photoionization model T2002 constrained the column density (\lNHI = 15.85
\cm2), $b$-value = 16.1 \kms,
metallicity ([C/H]=-1.2 and [Si/C]=+0.2 in solar units) and the density 
(log $n_H$= -2.8 cm$^{-3}$); these values determine the line-of-sight
size of this absorber to be 70 pc.

We discuss the hydrogen and metal absorption lines at Virgo distances 
($cz$=800-2400 \kms) that we have identified in the available
STIS echelle and FUSE spectra of RX J1230.8+0115 (\S 2). In \S 3 we discuss in 
detail our inferences concerning the N$_{HI}$, metallicity and ionization 
parameter for the $cz$=1699 \kms\ absorption complex based upon the HST and FUSE 
data, and the assumption of photoionization equilibrium. 
In \S 4 we discuss the galaxy distribution around these absorbers; in \S 5 we
discuss the inferences based upon the coincident absorption line pairs in these 
two spectra. \S 6 contains both a summary of our
observational results and a discussion of the model that best fits all of the 
information currently available on the low ionization, coincident velocity 
absorbers in these two sight lines.
Throughout this paper we use $H_0$=70 \kms\ Mpc$^{-1}$ and determine distances 
assuming a pure Hubble flow; the use of the Tonry et al. (2000) Virgocentric
infall model reduces distances in this vicinity by $\sim$10\% for galaxies with
$cz = 1500 - 1800$ \kms. We cannot rule out that redshifts yield a
double-valued distance solution at this local Supercluster location (Tonry et
al. 2000).

\section{UV Spectroscopy of RX J1230.8+0115} 

We describe the HST and FUSE spectroscopy of RX J1230.8+0115 at
recession velocities which correspond to Virgo Cluster distances, with 
primary emphasis on a very strong, two-component absorber at $\sim$1700 \kms. 
While the emission line redshift of this AGN ($z$=0.117) 
allows detection of absorbers at higher
redshifts than Virgo, they will not be discussed here. First we
describe the \lya\ and metal-line absorptions seen in an echelle mode
spectrum obtained with STIS, then show
the higher-order Lyman series lines for the strongest two absorbers at
Virgo distances obtained with FUSE. An analysis of a
GHRS first order grating spectrum of  RX J1230.8+0115 was presented by Penton, 
Stocke \& Shull (2000a), in which 5 of the 6 absorbers tabulated here were 
detected.

\subsection{\lya\ Absorbers}

Observations of RX J1230.8+0115 were obtained with the E140M echelle mode of STIS
using the 0.2\arcs$\times$0.06\arcs\ aperture on January 1 and January 13, 1999
(M. Rauch, PI). The resolution of the spectrum is
7 \kms\ over the spectral range of 1150 to 1710 \AA. The data reduction of the 
spectrum was done using the standard STIS extraction
procedures. This reduction package returns heliocentric velocities
which are estimated to be accurate to $\pm$ 4 \kms\ (STIS data handbook). The 
RX J1230.8+0115 spectrum has considerably lower signal-to-noise (SNR) 
than the 3C~273 spectrum of \citet{Heap:02}
because it is $\sim$ 10 times fainter. Nevertheless, the
Virgo distance \lya\ absorbers are strong enough to be easily detected (see also
Penton, Stocke, \& Shull 2000a).


Figure \ref{fig:lya} shows the region of \lya\ absorption for Virgo velocities
($cz <$ 2500 \kms) in the RX J1230.8+0115 sight line. We have binned these data 
by 3 pixels, giving a pixel width of 9.7 \kms\ and removed a continuum that is
the best by-eye fit to the damped \lya\ profile. We get a best-fit value for
Galactic N$_H = 1.6\times10^{20}$ \cm2\ scaled to a continuum. This fit is shown 
in Figure \ref{fig:lya} (solid line). We also show a spline fit
(dashed line) that was used to estimate line significance between 1218 and 1221
\AA\ because the Galactic damped \lya\ fit appears to underestimate the true 
continuum in this region indicating that the AGN continuum must be rising 
blueward of 1221 \AA. Also shown are the errors in the flux measurements. The 
absorption lines were fit with Voigt profiles using VPFIT \citep{carswell:02}. 
We identify 5 lines (one of which has 2 components) at 4-$\sigma$ or greater
significance, and one line at greater than 3.5-$\sigma$ significance. Because 
there are no Galactic metal lines and no metal lines
related to the associated absorbers, we identify all of the absorptions as 
intergalactic \lya. Table 1 contains the following information about these
absorbers: (1) central wavelength of the Voigt profile fit in \AA, (2) central 
velocity of the fit (heliocentric, non-relativistic, in \kms) (3) the redshift 
and (4) its associated error, (5) the $b$-value and (6) its associated error in 
\kms, (7) the log of the H I column density and (8) its associated error in \cm2
-- these errors include a continuum placement error derived by adding and 
subtracting one-third of the rms from the continuum, (9) the rest frame 
equivalent width (EW) of the data (not of the Voigt profile fit) measured in 
m\AA, (10) the 4-$\sigma$ EW limit integrated over the same velocity width for
which the EW in column 9 was calculated (typically 3 -- 4 times the $b$-value in
column 5). All of these values, except for the 
lines at 1685 and 1721 \kms, are single component fits determined from the 
\lya\ absorptions alone. For these two components (1685 and 1721 \kms) we have
fixed the redshifts (based upon the observed metal line redshifts discussed 
in the next section) and the column densities (at \lNHI = 16.2
and 16.6 \cm2\ as discussed in \S 2.4), which is why there are no quoted errors.

\centerline{\epsfxsize=0.9\hsize{\epsfbox{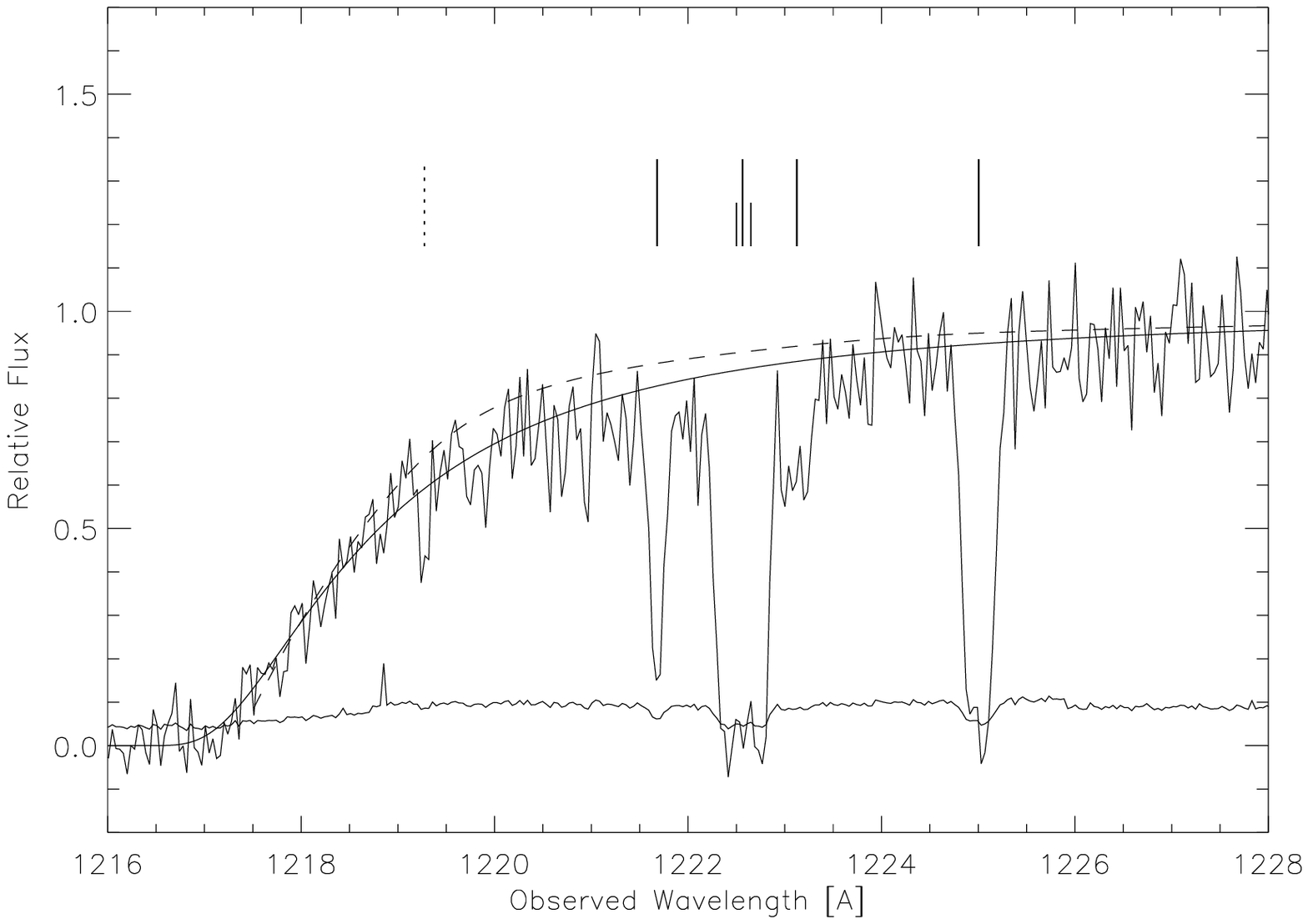}}}
\vspace{0.1in}
\figcaption{\label{fig:lya} The region of \lya\ absorption for Virgo velocities in the 
RX J1230.8+0115 sight line. The data have been binned by three pixels and the 
continuum has been fitted with a damped \lya\ profile (solid line; see text) and
a spline fit (dashed line). The spline fit is used to fit the region between
1218 and 1221 \AA\ since Galactic damped \lya\ fit appears to underestimate the 
continuum slightly in this region, indicating that the AGN
continuum is rising blueward of 1221\AA. The vertical marks above the spectrum 
indicate the $\geq$4$\sigma$ absorption lines detected 
(see Table 1). The dotted vertical mark indicates the $\geq$3$\sigma$ absorption
line that has been detected in this region. The two smaller ticks represent the 
positions of the metal line components in the strong absorber blend. The lower 
curve shows the errors in the flux as a function of wavelength. \\}

The 4-$\sigma$ EW limits given in Table 1 indicate that
all of these absorption-line detections are stronger than 4-$\sigma$ except for
the line at 1219.27 \AA\ which is 3.5-$\sigma$. Our 
strongest absorbers, the $z = 0.005668$ (1699 \kms) blend and 
the $z = 0.007680$ (2302 \kms) absorption, are both highly 
saturated, which makes it much more difficult to obtain accurate measurements of
column densities and $b$-values. In fact, for the 1699 \kms\ blend, we have run 
an entire grid
of H I column density models and found little discrimination based upon the
\lya\ fits alone over the range 10$^{15.0-17.6}$ \cm2\ (reduced $\chi^{2}$
values of 1.1-2.1 over this range). Because the \lya\ data are
insufficient for this task, we discuss further constraints on the H I column 
density in the $z = 0.005668$ complex in \S 2.4. 

\subsection{Metal Lines Associated with the RX J1230.8+0115 Absorbers}

We have examined the RX J1230.8+0115 STIS echelle spectrum for metal lines 
associated with the
H I absorptions discussed above and detect metal-line absorption in only the
$z = 0.005668$ complex. Two component metal lines of C~II,
C~IV, Si~II, Si~III, and Si~IV are found associated with this complex. We used
the FUSE data (see \S 2.3) to search the region of C~III absorption at this
redshift, but we do not fit the C~III absorption because the spectrum is 
noisy and the potential C~III absorptions are blended with Galactic H$_2$. We 
also do not fit the Si~III absorptions in the STIS echelle data because they are
far down on the damping-wing of Galactic \lya\ and, therefore in a region of
very low S/N. The local continuum around the metal
lines was determined with a linear least squares fit, as the regions were
determined to be flat. The
metal lines were then fit using VPFIT as for the \lya\ lines. 
Figure \ref{fig:1666lines} shows these metal lines and the VPFIT results are 
summarized in Table 2. Except for column (1), which lists the identified ion, 
the columns have the same descriptors as Table~1. The final column gives
4-$\sigma$ limits which are also appropriate for metal line non-detections in
the other \lya\ absorbers listed in Table 1. The statistical errors on 
the $b$-values listed in column (6) are errors returned on the specific line 
profile fits in Figure 2.



\centerline{\epsfxsize=0.9\hsize{\epsfbox{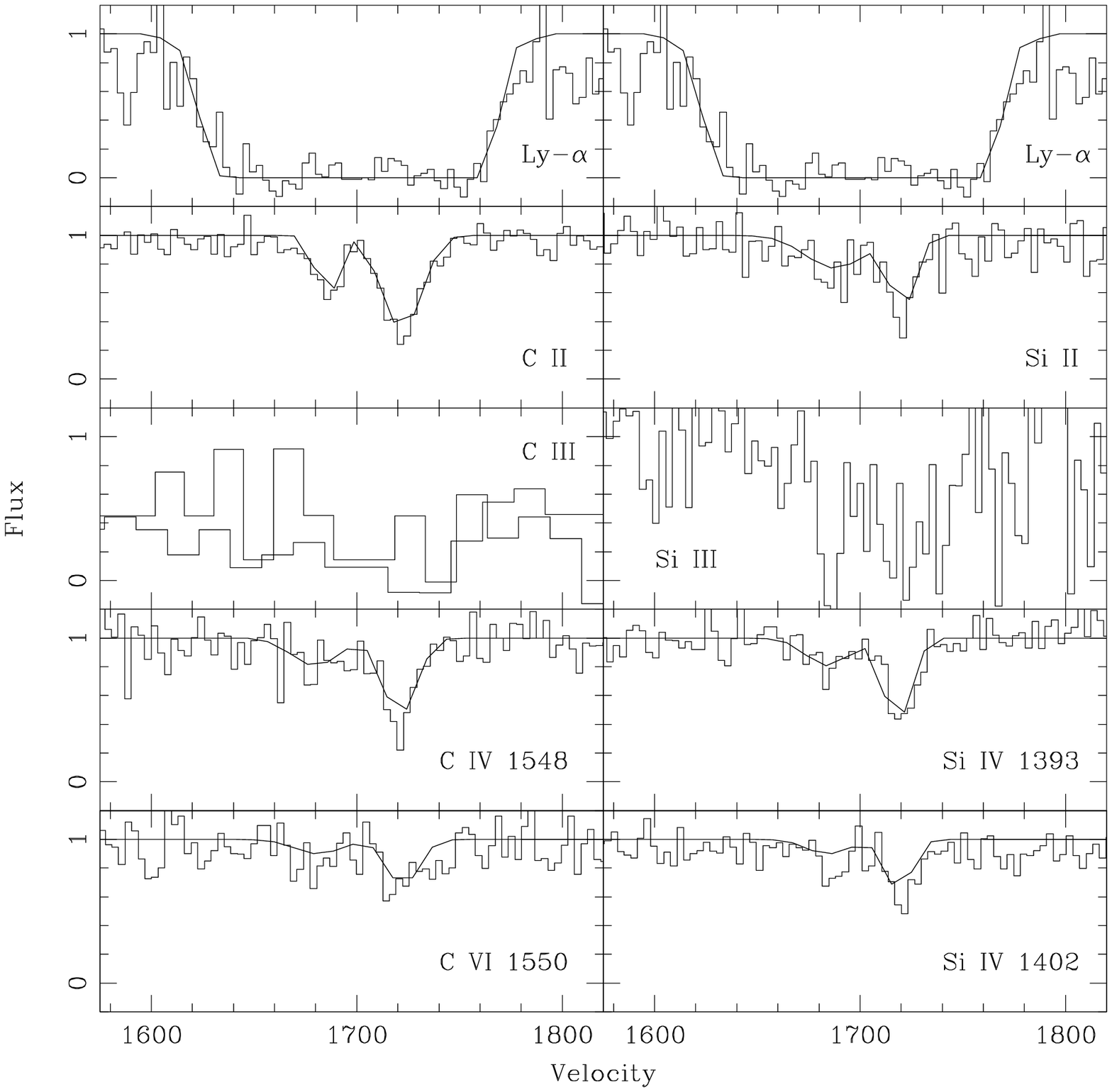}}}
\vspace{0.1in}
\figcaption{\label{fig:1666lines} The metal lines associated with the 
$z = 0.005668$ absorber. We see two
component lines corresponding to C~II, C~IV, Si~II, Si~III, Si~IV. The region of
C~III in the FUSE data has been shown, but it has not been fit because it is
blended with Galactic H$_2$. The two histograms in the C~III plot show the SIC1B 
and SIC2A data separately. There is no fit to the Si~III data because Si~III is 
in a very low S/N region on the damping wing of Galactic \lya\ making a fit to 
this line highly unreliable. 
However, we see the same two-component absorption in the Si~III and possibly 
C~III as in the other ions. The STIS data have been displayed with no binning to
show the full resolution noise level of the spectra. \\} 

In all cases where metal lines are detected, two components are detected well
within the broad, saturated, \lya\ profile shown at the top of Figure
\ref{fig:1666lines}. Because the velocities of the two components are
consistent between species and ionization states, we have used the velocity of
the strong C~II components to constrain the velocities of the 1685 and 1721 \kms\
components in the Lyman line analysis. The consistency of the column density
ratios and the line profile shapes provides an excellent indication that the gas
is photoionized and has a simple kinematic structure. A substantial contribution
from collisionally-ionized gas is unlikely.

The metal lines all have resolved line profiles that provide a measurement of
the total $b$-value. These line widths are inconsistent with being due to pure
thermal broadening since photoionization models predict temperatures below 20,000 
\deg K while a thermal broadening interpretation of the line widths implies 
temperatures of $\sim$90,000 \deg K.
However, the line widths of the 1720 \kms\ line (the signal-to-noise of the 1685 
\kms\ line is too low to provide reliable b-value constraints) are consistent
within 2-$\sigma$ with the same amount of bulk or turbulent motion for all of 
the ionization states.

\subsection{FUSE Observations of Higher-Order Lyman Lines}

RX J1230.8+0115 was only observed in three brief ``snapshots'' totaling 4031 
seconds with FUSE. Nevertheless, we examined the co-added snapshots
to better constrain the H I column density of the $z = 0.005668$ complex. The 
spectrum has a SNR of 1.1 for the SIC1B channel and 1.5 for the SIC2A
channel. We have binned the data by 8 pixels for a resolution of 16 \kms\ per 
pixel. The data were processed using CALFUSE version 2.1.6. 


\centerline{\epsfxsize=0.95\hsize{\epsfbox{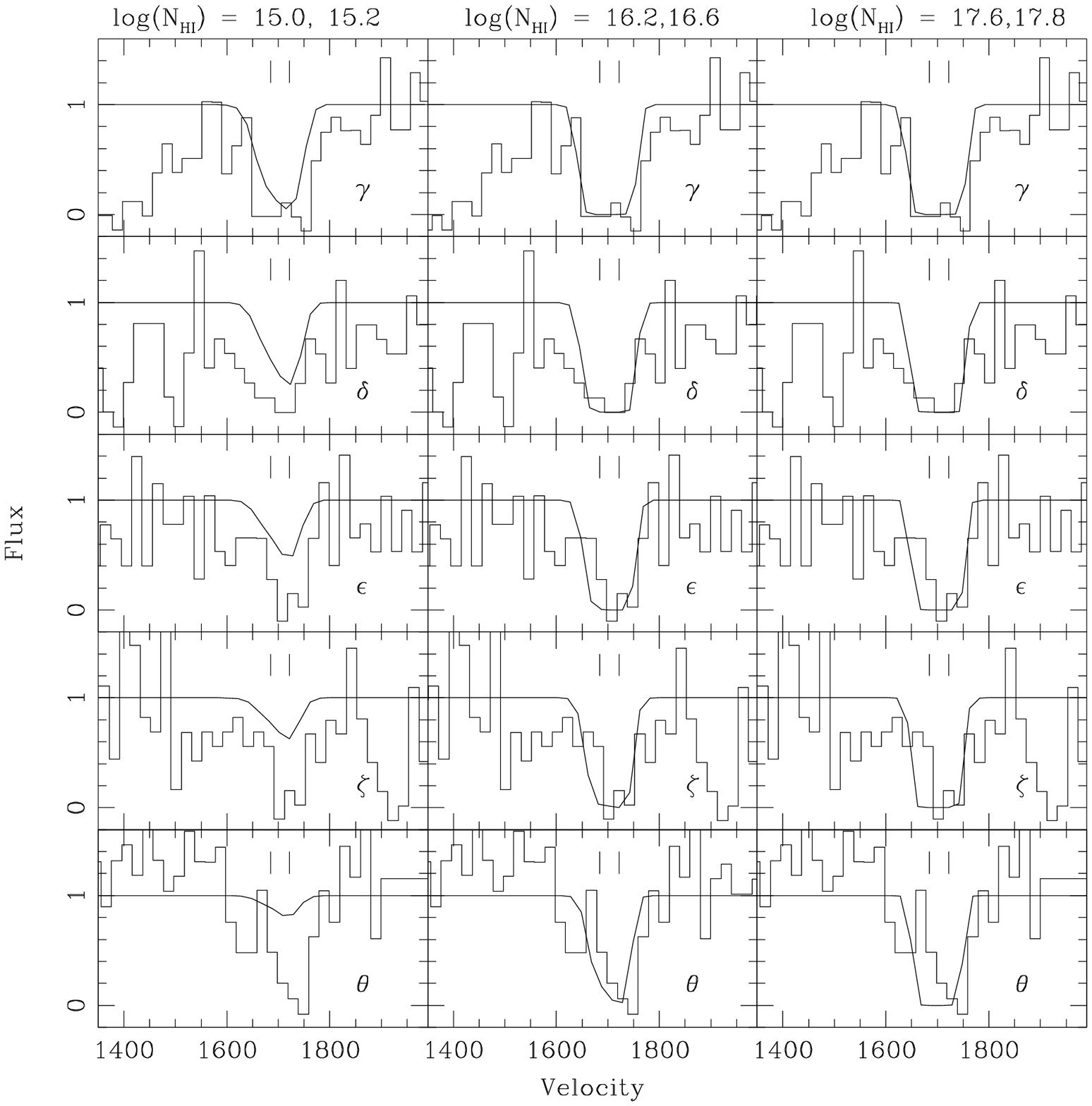}}}
\vspace{0.1in}
\figcaption{\label{fig:fusefits} Fits to the FUSE data on the Lyman series lines 
associated with the 
$z = 0.005668$ absorber, spanning a plausible range of column densities based on 
fits to the \lya\ line. The central fit is for a metallicity of 0.06 
Z$_\odot$ for each component. These data indicate that fits as low as \lNHI = 15 
\cm2\ are excluded and that fits with \lNHI $\geq$ 16 \cm2\ are probable. \\}

Figure \ref{fig:fusefits} shows the portion of the Lyman series detected for 
this absorption complex. Lyman-$\beta$ has not been included because it is 
severely blended 
with a Galactic O~VI line and Lyman-$\eta$ has also been omitted because it is 
blended with Galactic Lyman-$\epsilon$. Various fits are shown spanning a
plausible range in H I column density bounded by acceptable fits to the \lya\ 
line which range from \lNHI = 15 -- 17.6 \cm2\ for the lower column density
component. For all of these fits we have assumed that the two components, fixed 
at the metal line velocities, have
comparable metallicities and, therefore, roughly consistent relative H~I column
densities. The Lyman-limit does not place a significant
constraint on the column density due to confusion with a plethora of Galactic
lines in the 915-920\AA\ range. Figure \ref{fig:fusefits} shows that 
even with the FUSE data, it is very difficult to distinguish between these 
column densities. It does appear that the column density is
larger than \lNHI = 15.2 and 15.0 \cm2\ for the two components, and probably larger
than 10$^{16}$ \cm2\ in both components. An assumption that the metallicity is 
$\sim 6$\% solar as in the T2002 absorber, implies \lNHI = 16.2, 16.6 \cm2\
(see \S 3) is not a bad one. However, the FUSE spectrum is inadequate to
constrain the column densities at the high end.

\subsection{Observational Constraints on the Column Densities of the H I 
Absorptions at $z = 0.005668$}

The saturation of the $z = 0.005668$ complex and the poor quality of the FUSE 
spectrum make it difficult to determine 
accurate column densities, so we list our best-fits and constraints for
these values, in order of most to least secure. 

\begin{itemize}
 
\item{We place a firm lower limit on the H I column density in this line-of-sight
by integrating the optical depth pixel by pixel within the line 
\citep{savage:91} using the equation:

\begin{equation}
{\rm N}_{\tau_v} = {{3.767\times 10^{14}}\over{f\lambda_0}}\,\sum_{i=1}^n
\ln\biggl[{{I_c(v_i)}\over{I(v_i)}}\biggr]\,{\rm d}v_i,
\label{eq:app_col}
\end{equation}

where $I(v_i)$ is the observed flux and $I_c(v_i)$ is the estimated 
continuum. If the observed flux is less than the rms in the flux, we assign the
flux to be equal to the rms for that pixel. $\lambda_0$ = 1215.67\AA, the rest
wavelength for the \lya\ transition, and f = 0.416 is the oscillator strength.  
The lower limit to the column density from this method is 3$\times 10^{14}$
\cm2.}

\item{The consistency of the metal line column density ratios argues for a
photoionization model while the lack of variation in the line profiles for 
different ions and species argues for a single phase for the gas.}

\item{While it is difficult to use the FUSE data to place accurate constraints on 
these two absorbers, it seems probable that the Ly-$\epsilon$, Ly-$\zeta$, and
Ly-$\theta$ absorptions requires \lNHI $>$ 16.0 \cm2.}

\item{We have used the redshifts of the C~II absorption lines in this complex 
to fix the positions of the H~I components. For the \lya\ saturated absorption
complex, the best-fit ($\chi_\nu ^2 = 1.1$) from VPFIT is found when \lNHI = 15.0 
\cm2\ for the $z = 0.00562$ component and \lNHI = 15.2 \cm2\ for the $z =
0.00574$ component. But these column densities are not consistent with the strong,
albeit noisy, higher Lyman series absorption seen with FUSE. Further, as we
discuss in \S 3, if the gas is photoionized such low values of \NHI\ imply
metallicities of order solar to account for the C~II and C~IV column densities.
Metallicities greater than 0.3 solar are unlikely, as this corresponds to the
metallicity of the intracluster medium at the center of the Virgo 
Cluster as derived from the X-ray observations \citep{mushotzky:97}. By
comparison, the metallicity found in the 3C~273
strong absorber at this velocity is 6\% solar (T2002). 
While we can derive acceptable \lya\ line fits for all column densities tried up 
to \lNHI = 17.8 \cm2\ for the $z = 0.00562$ component and \lNHI = 18.0 \cm2\ for
the $z = 0.00574$ component, for column densities exceeding 10$^{17}$ \cm2\ the
$b$-values derived to fit the \lya\ equivalent widths are unrealistically low 
(8-14 \kms) for these absorbers. 
While such low measured values are possible if these absorbers have no turbulent 
motion \citep{dave:01}, 
the presence of C~IV and Si~IV at comparable strength to C~II and Si~II make such
low values very unlikely; e.g., the 3C~273 absorber studied by T2002 has no C~IV and $b$=
16.1 \kms. We conclude that the most likely column density range that fits all the data is
N$_{H I}$=10$^{16-17}$ \cm2.}   

\item{If we assume that the metallicities of the components are the same
as for the T2002 absorber in the 3C~273 sight line, we derive 
column densities of 1.6 $\times 10^{16}$ \cm2\ and 4.0 $\times 10^{16}$ 
\cm2.  The \lya\ fit has a $\chi_\nu ^2$ = 2.04 (probability = 0.005), 
insignificantly different from the nominal best fits discussed above. These are 
the values shown in the middle panel of Figure 3 and are clearly consistent with
the FUSE data. These values also have reasonable, although somewhat low,
$b$-values for low-$z$ \lya\ absorbers of 17 and 16 \kms\ respectively.}

\end{itemize}

In conclusion, while we cannot prove based upon the data alone that the column
density of these two absorbers is \lNHI = 16.2 and 16.6 \cm2\ and that the
metallicities are a few percent of solar, these values are the most probable 
for this absorption complex. Our ``best guess'' model has \lNHI = 16.2,
16.6 \cm2\ derived from the assumption that $Z$=0.06 $Z_\odot$, as in the T2002
absorber, for the two components. 

\section{A Photoionization Model of the $z = 0.005668$ Absorber Complex}

Because the highly saturated \lya\ absorption 
profile and the poor quality of the FUSE spectrum make it difficult to constrain 
the H~I column density precisely,
we use the metal line ratios to determine the properties of the gas
before returning to analyze the H~I absorption informed by the metal-line data.
As summarized in Table 2, the metal lines are associated with two velocity 
components, one at $v \sim 1685$ \kms, and one at
$v \sim 1721$ \kms. These velocity components show no differences in structure
as a function of species or ionization state suggesting that the gas is in a 
single phase. Several lines of argument favor a photoionization model
for the gas: (1) for the $v \sim 1721$ kms$^{-1}$ component, limits on the 
line width constrain the temperature of the gas to be less than 105,000 \deg K;  
(2) the relative Si II and C II column densities of both components are 
compatible with equal ionization fractions (after accounting for the relative 
abundances of Si and C); (3) Purely collisional ionization is inconsistent with
the Si II/C II ratio for temperatures above 20,000 \deg K (Sutherland \& Dopita 
1993). However, at temperatures below 20,000 \deg K extremely little C IV or Si 
IV would be observed. 

Thus, we consider the properties of the gas at each 
velocity in turn assuming: (1) all observed ionization stages of carbon and 
silicon are associated with the same gas, which is compatible with the ratio of 
column densities, as we will show; (2) the gas is primarily photoionized, which 
is consistent with the predominance of lower ionization stages for carbon and
silicon and with the consistency of the column density ratios; (3) the gas is 
approximated as a plane parallel slab illuminated by a
quasar-dominated spectrum \citep{shull:99,haardt:96} with an intrinsic power 
law spectrum of spectral energy index $\alpha = 1.8$.

To model the physical properties of the gas we have constructed a large grid of 
photoionization models using the photoionization code Cloudy \citep{ferland:96}. 
The calculated column densities of C~II, C~IV, Si~II, Si~III, and
Si~IV depend primarily on the metallicity, H~I column density, and ionization 
parameter $U = n_\gamma/n_H$. Relative to these parameters, the uncertainty in 
the spectral shape of the extragalactic background radiation contributes a
negligible difference to the models.

At this point, we examine the dependence of
the N(C~II)/N(C~IV), and N(Si~II)/N(Si~IV) column 
density ratios on the ionization parameter.  For
a given metallicity and H~I column density, we can
plot these ion ratios versus U, as shown in Figure \ref{fig:modcolU}. In this
figure we assume the metallicity to be $(C/H) = 0.06$ solar 
(corresponding to \lNHI = 16.2 and 16.6 for the 1685 and 1721 \kms\ absorbers
respectively), our ``best guess'' model based upon the constraints described in 
\S 2.4. Finally, closest agreement with the observed column densities
of N(Si~II) and N(Si~IV) is obtained by assuming log (Si/C) $\approx +0.1$, 
which is also compatible with the conclusions of T2002 for the absorber toward 
3C~273. The range in ionization parameter consistent with the carbon line 
observations is $ -2.89 < \log U < -2.65$ for the gas at $v \approx 1685$ km/s
and $-2.85 < \log U < -2.76$ for the gas at $v \approx 1721$ km/s. This range 
is also consistent with the N(Si~II)/N(Si~IV) column density ratio, so our one 
phase approximation likely remains a reasonable one. These ranges indicate the 
uncertainty in U, for a {\it given} metallicity or H~I column density.


To explore the sensitivity of U to the H~I column density and metallicity, we 
step through column density within the range, \lNHI = 14.7 -- 17.5 \cm2,
corresponding to a 
metallicity in the range 6 -- 0.005 $Z_\odot$ such that the absolute 
column density of each metal ion is always consistent with the values in Table 2. 
The range in U is fairly insensitive to \NHI\ and $Z$; the expanded 
range becomes $ -2.9 < \log U < -2.3$ and $ -2.9 < \log U < -2.4$
for the gas at $v \sim$ 1685 and 1721 \kms\ respectively over this range of
\lNHI.

For a single-phase model, as suggested by the simplicity of the metal line
profiles, the allowed range in ionization parameter may be 
translated into a range in density if the intensity of the extragalactic
ionizing background is known. \citet{shull:99} estimate a mean intensity 
of $J_\nu \sim 1 \times 10^{-23}$ ergs~s$^{-1}$\cm2 ster$^{-1}$Hz$^{-1}$ 
at one Rydberg by summing the contribution from quasars, which is
consistent with $J_\nu$ calculated using more direct methods such as the 
low-$z$ proximity effect, truncation of H~I disks, and limits on H-$\alpha$ 
from extragalactic clouds \citep{bechtold:94, shull:99, maloney:93, Weymann:02, 
Donahue:95}. The implied range in hydrogen density for these absorbers is 
$n_H$ = 1 -- $5 \times 10^{-4}$ and 1 -- 3 $\times 10^{-4}$ cm$^{-3}$ for the 
$v \sim$ 1685 and 1721 \kms\ components respectively. These densities 
correspond to overdensities of $\delta \sim$ 600 -- 3000 and 600 -- 1800 for 
the two absorbers in units of the mean baryon density at $z = 0$. These
densities are an order of magnitude lower than those inferred for the gas
towards 3C~273 (T2002).

\centerline{\epsfxsize=0.9\hsize{\epsfbox{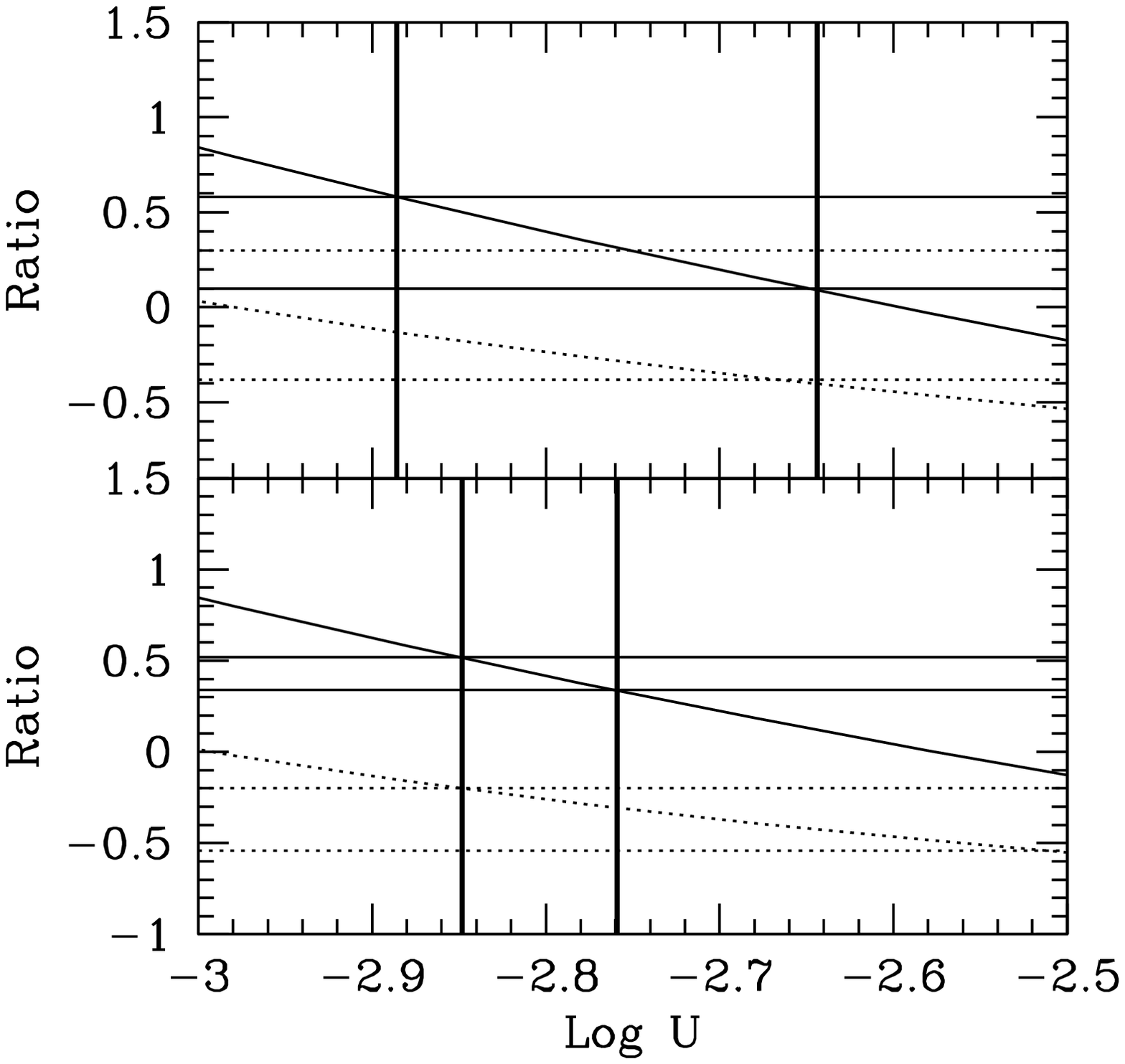}}}
\vspace{0.1in}
\figcaption{\label{fig:modcolU} Calculated column density ratios versus 
ionization parameter U for models approximating the absorber at 1685 \kms\
(upper panel) and at 1721 \kms\ (lower panel).  Solid curves
indicate the changing N(C~II)/N(C~IV) ratio and dotted curves
indicate the changing N(Si~II)/N(Si~IV) ratio.  Solid [dotted]
horizontal lines indicate the 1$\sigma$ uncertainty about
the measured values of N(C~II)/N(C~IV) [N(Si~II)/N(Si~IV)].
The thick vertical lines indicate the range in Log U consistent
with the observations of the N(C~II)/N(C~IV) ratio.  As shown, the range
in Log U is also consistent with the N(Si~II)/N(Si~IV) ratios
in the 1721 \kms\ absorber.  The range in Log U in the 1685 \kms\
absorber is marginally inconsistent with the
N(Si~II)/N(Si~IV) ratios.  We have chosen to use the more
secure N(C~IV)/N(C~II) measurements to set the acceptable
range in the ionization parameter. \\} 


The extent of the gas associated with the absorption, over the possible range of 
\NHI\ or metallicity is $0.8 - 50$ kpc. However, the requirement that our models
reproduce the absolute C~II and C~IV column densities, as well as their
ratios, means that the \NHI\ and metallicity for successful models are highly
correlated, as shown in Figure 5.  Since the line of sight extent of the gas in 
our models is derived by dividing \NHI\ by the mean hydrogen density, and since
the range in mean hydrogen density compatible with
line ratios varies slowly with \NHI\, there is also
a good correlation of extent with assumed metallicity.
Figure \ref{fig:extent} shows the line of sight extent
of the gas as a function of the assumed metallicity. Very large extents are
possible only by invoking $Z \leq 10^{-2} Z_\odot$;
these low metallicities also require \lNHI $ >$ 17 \cm2,
which yield unrealistically low $b$-values (see \S 2.4).
For the probable lower limit on \lNHI $\ge$ 16.0 \cm2\ suggested by the
FUSE detection of higher order Lyman lines,
the metallicity of these structures must be \lapprox 10 -- 20\% solar,
in agreement with their location well away from the Virgo core.

\centerline{\epsfxsize=0.9\hsize{\epsfbox{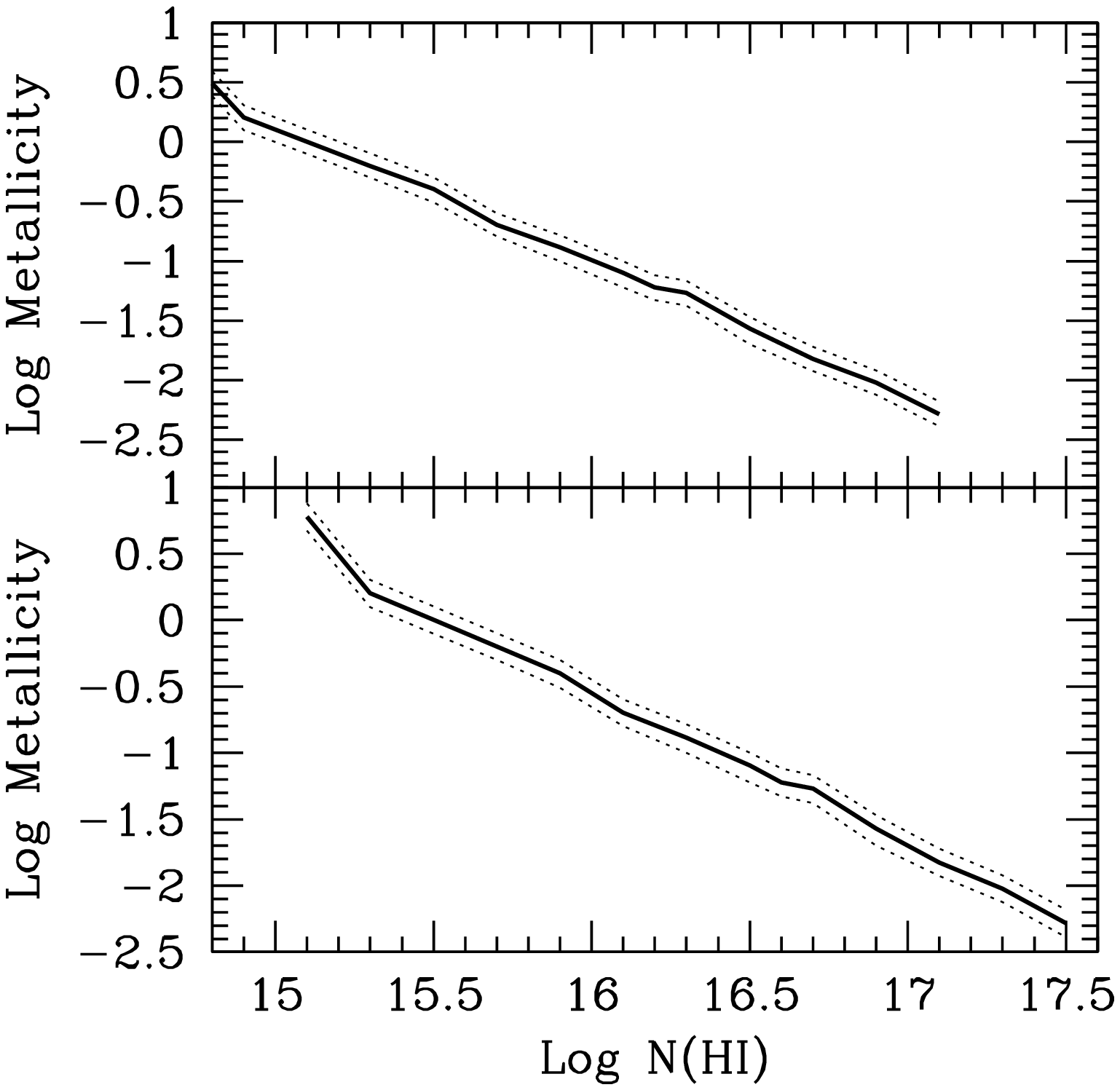}}}
\vspace{0.1in}
\figcaption{\label{fig:metals} The range in metallicity consistent with
observed column densities for
photoionization models approximating the absorber at 1685 km/s
(upper panel) and at 1721 km/s (lower panel). The dotted curves show the model
uncertainty in this relation. \\}


\centerline{\epsfxsize=0.9\hsize{\epsfbox{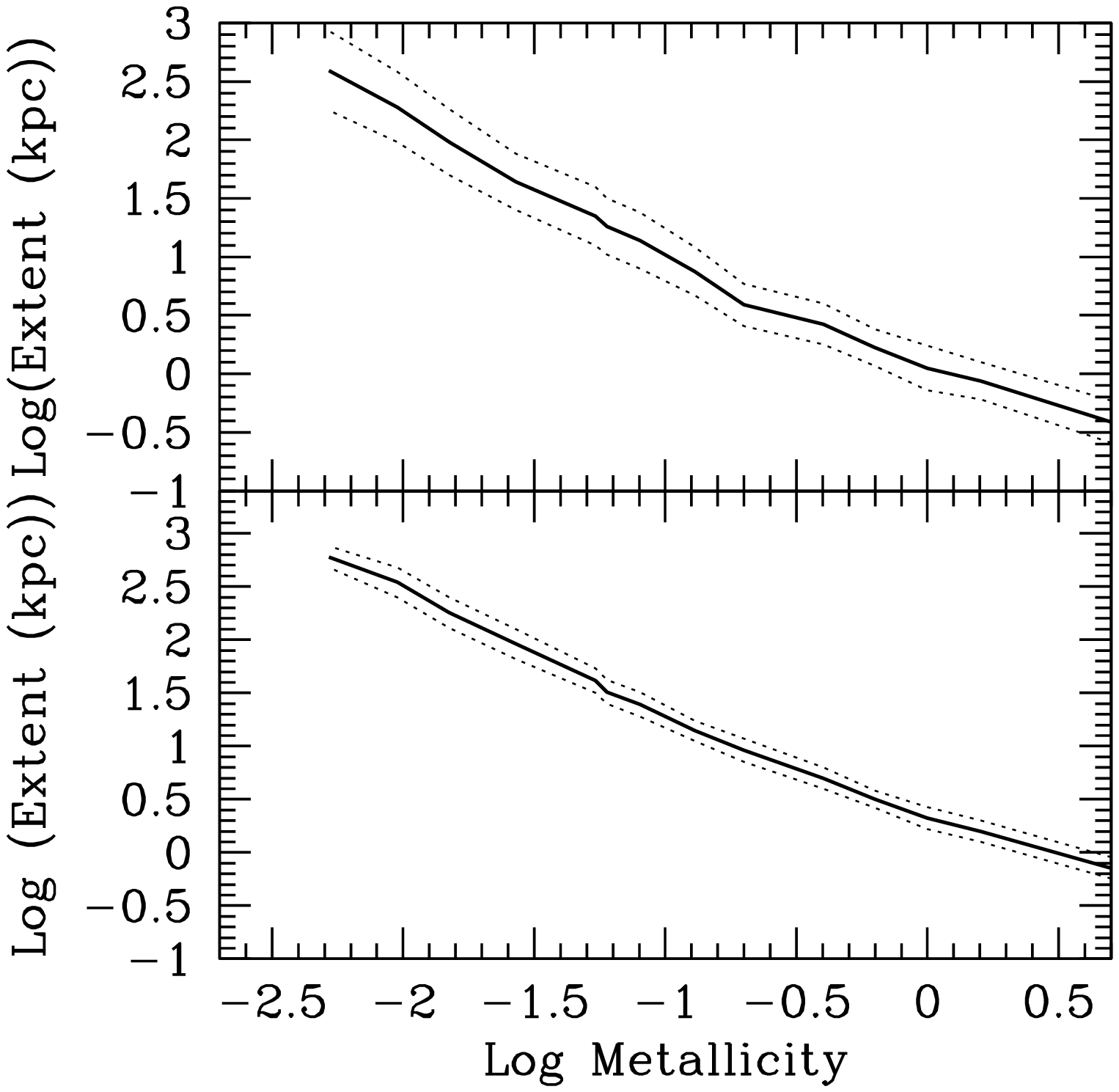}}}
\vspace{0.1in}
\figcaption{\label{fig:extent} Calculated line-of-sight extent versus metallicity
parameter for models approximating the absorber at 1685 \kms\
(upper panel) and at 1721 \kms\ (lower panel).  The dotted lines represent the
model uncertainty in this relation. \\}

For our ``best guess'' model, \lNHI = 16.2 and 16.6 \cm2, derived from the 
assumption of $\sim$ 6\% solar abundances 
(consistent with the nearby 3C~273 absorber) we find sizes along the line 
of sight of $\sim$10 -- 30 kpc. The T2002 analysis of the 3C~273 absorber at the
same velocity infers an even smaller line-of-sight depth of only 
70 pc. We cannot rule out that the
column densities are an order of magnitude higher
than our ``best guess,'' with corresponding metallicities of
$\leq$0.01$Z_\odot$, and much larger line of sight extents.
However, the observations together with those
analyzed in T2002 suggest that a very thin ($\sim$ 10 kpc) ``ribbon
or sheet'' of photoionized gas with abundances
of a few percent of solar extends for
several hundreds of kpc through this region on the outskirts of the Virgo
Cluster or that there are several thin gaseous structures lying along a large 
scale filament in this region. Given the expected velocity dispersion of 
galaxies in the outskirts of a cluster like Virgo (100-300 \kms), it is unlikely 
that a single thin
``ribbon'' would be dynamically stable for the length of time required to form 
it (\gapprox 10$^9$ yrs).

\section{Galaxy Identification in the Vicinity of RX J1230.8+0115 and 3C~273}

Our knowledge of galaxies near the RX J1230.8+0115 sight line at Virgo Cluster
velocities comes from four sources: (1) an
augmentation to the Center for Astrophysics (CfA) redshift survey 
\citep*{grogin:98}, (2) a pointed survey using Las Campanas Observatory (LCO) to
obtain redshifts around 3C~273 \citep{morris:93}, (3) two deep pointed 
surveys using multi-object spectroscopy from the Wisconsin-Indiana-Yale-NOAO 
(WIYN) 3.5m\footnote{The WIYN Observatory is a joint facility of the University 
of Wisconsin -- Madison, Indiana University, Yale University, and the National 
Optical Astronomy Observatories.} and from the Las Campanas Observatory (LCO) Du 
Pont 2.5 telescopes \citep{stocke:03}, and (4) the Sloan Digital Sky Survey
(SDSS) early data release (EDR, Stoughton et al. 2002). The pointed galaxy 
surveys were designed to investigate the relationship between the \lya\ forest 
absorption systems along these sight lines and the galaxies.

\citet{grogin:98} obtained spectroscopic data to augment the CfA redshift 
survey in the 3C~273 region so that the survey would be virtually complete 
down to $m_b \leq$ 15.7 (M$_{B} \leq$ -16.2 at 24.3 Mpc for $cz$= 1700 \kms) in 
a region bounded 
by 11.5$^h \leq \alpha \leq 13.5^h$ and -3.5\deg $ \leq \delta \leq$ +8.5\deg . 
This region includes both the 3C~273 and RX J1230.8+0115 sight lines and extends
several Mpc around these sight lines at Virgo distances. The SDSS EDR
spectroscopic survey includes a strip around the north Galactic cap at $\delta
= 0$\deg\ that extends up to $\delta = 1.3$\deg, covering the declination of
RX J1230.8+0115 but not the declination of 3C~273. The nominal completeness
limit for this spectroscopic survey is r~$<$~17.8, at least two magnitudes 
fainter than the augmented CfA survey in this region.

\citet{morris:93} also conducted a galaxy redshift survey around 3C~273 for 
which they claimed completeness to m$_B\sim$19 over a region with a 60\arcm\
radius in $RA$ and 40\arcm\ radius in $DEC$ from the quasar.  While RX J1230.8+0115 is 
just beyond the southern edge of this survey region, it includes 
galaxies up to $\sim$ 700\h70\ kpc to the north and west of  RX J1230.8+0115. 

More recently, \citet{stocke:03} have conducted a multi-object spectroscopy survey 
in the region, including: (1) an LCO slit-mask spectroscopy survey 
around RX J1230.8+0115; (2) an LCO slit-mask survey around 3C~273 to augment the 
\citet{morris:93} observations; and (3) a 
WIYN/HYDRA fiber-fed spectroscopic survey in the region between 3C~273 and 
RX J1230.8+0115 
centered at $\alpha = 12.5^h\ \delta = +1.67$\deg. A preliminary discussion of 
the observing setups, data reduction, and analysis of the LCO and WIYN data can 
be found in \citet{mclin:02}. A more complete discussion of the techniques and 
results for all sight lines observed will be presented later \citep{stocke:03}.
We review what has been done so far in the 3C~273/ RX J1230.8+0115 region by
this on-going survey here:

(1) The LCO observations surrounding RX J1230.8+0115 consist of spectra for 43 
galaxies, or $\sim$ 78\% of the galaxies down to $m_B \sim$ 20 (corresponding 
to $M_B \leq$ -12 at the distance of the $z = 0.005668$ complex) out to a radius 
of 5\arcm\ (36 h$^{-1}$ kpc at this distance) from the target and 21\% of the galaxies 
out to 15\arcm. No galaxies were found at the redshifts of any of the Virgo 
Cluster absorbers, although several galaxies associated with absorbers behind 
Virgo were found.

(2) The LCO slit mask spectroscopy around 3C~273 is 100\% complete down to 
m$_B\sim$19.5 out to a 10\arcm\ radius. We obtained redshifts for 5 galaxies
that were missed by \citet{morris:93}, probably due to a 
classification error made by the automated galaxy-finding software employed  
\citep[see][]{stocke:03} despite their claim of 100\% completeness over a large
region. Seventy-seven galaxies fainter than the \citet{morris:93} magnitude limit 
were also observed. 

One of the five galaxies missed by \citet{morris:93} is a relatively bright 
(M$_B = -14.5$) galaxy 71\h70\ kpc from 3C~273 (360\h70\ kpc
from RX J1230.8+0115) at the redshift of the strong H~I absorber
($cz_{abs}$=1589 \kms) analyzed by T2002. This galaxy has the 
spectroscopic signatures of a ``post-starburst,'' a rare galaxy type in the local 
Universe. The
starburst phase might be expected to create a substantial wind that could be
responsible for a high-column density, low-ionization absorber like the one
present in the 3C~273 sight line \citep{stocke:03}. While this galaxy is too far 
from RX J1230.8+0115 to be related to the $z = 0.005668$ complex, the fact that 
this object was missed by \citet{morris:93} due to a classification error,
suggests that the region around RX J1230.8+0115 may not have been surveyed 
well enough to declare that similar luminosity galaxies have not been missed.
No other galaxies were found in these observations at Virgo distances.

(3) The WIYN/HYDRA observations concentrated on the region between RX 
J1230.8+0115 and 3C~273, 0.9$^\circ$ to the north. The fiber setup 
for these observations was centered at RA = 12:30:00 and  DEC = +01:40:00 
(J2000.0) and redshifts were obtained for 81\% of the galaxies within a 
20\arcm\ radius of this position. A total of 38 galaxies were observed 
successfully; only one galaxy was observed at both WIYN and LCO as we worked to 
prevent duplication as much as possible. Thus, in the region north of 
RX J1230.8+0115, towards 3C~273, our galaxy survey is $\sim$80\% complete
down to m$_B\leq$18.7. No galaxies were found at Virgo distances in the 
WIYN/HYDRA observations either.

The galaxy survey work should provide some confidence that this region 
of space has been scrutinized exceptionally well for galaxies. There are
17 galaxies with $cz\leq$ 2500 \kms\ now known in the region within 1 Mpc of
either sight line. The SDSS (Stoughton et al. 2002) and the \citet{morris:93} 
galaxy survey work is almost completely overlapping in the RX J1230.8+0115
region to an absolute magnitude limit of M$_B \simeq -14$. The CfA and Stocke et
al. (2003) survey work suggests that all
relatively bright galaxies (M$_B \le -16$) have been identified in the field and
that only extremely faint (M$_B \le -12$), extremely compact, or extremely low 
surface brightness galaxies could have been missed within 36\h70\ kpc
of the sight line. A summary of the 
nearest galaxy to each of the
absorbers is included in Table 3, which lists by column: (1) target sight line; 
(2) \lya\ absorber redshift (we list only the centroid of the absorption blend 
modeled here); (3) the heliocentric absorber velocity in \kms; (4) the velocity 
difference in the sense of ($V_{abs}$
- $V_{gal}$) in \kms; (5) the impact parameter in \h70\ Mpc; and (6) the name of
the nearest galaxy to each absorber. The relationship between the \lya\ absorbers
and seven of these galaxies has been discussed extensively by
\citet{penton:02}. The galaxy Leda 139866 is at 2329 \kms, just out
of the velocity range discussed by \citet{penton:02}, and the ``new galaxy" is the
post-starburst galaxy described briefly above. The galaxy labeled
[ISI96]1228+0116 is a very low surface brightness dwarf irregular (M$_B \approx
-13.6$; Impey et al. 1996).

Ten of the 17 galaxies near RX J1230.8+0115 and 3C~273 have recession velocities 
within $\pm$300 \kms\ of the $z = 0.005668$ complex, but all are at substantial 
impact parameters (280 -- 520\h70\ kpc). The nearest galaxy is an edge-on 
spiral of modest luminosity (CGCG 014-064; M$_B$ = -16.2) located 279\h70\
kpc away with a redshift nearly identical to that of one of the absorbers in this 
complex. If the galaxies responsible for absorption at large impact parameters
are expected to be bright, there is a much brighter, face-on spiral (NGC 4517A; 
M$_B$=-18.8)
a little further away, $\rho$= 365\h70\ kpc, with a larger velocity difference 
($cz_{gal}$ = 1530 \kms). On the other hand, it has been argued that LSB galaxies may
be responsible for much of the \lya\ absorption that we see. There is an LSB 
galaxy, discovered by \citet{impey:88}, in the field, but it is even further away 
in impact parameter (423\h70\ kpc) and velocity ($cz$ = 1473 \kms). However,
the velocity is similar to that of the absorber at $cz$=1482 \kms (see Table 1). 
Identification of this LSB galaxy with the 1482 \kms\ absorber is problematic 
since there are three other galaxies within $\pm 350$ \kms\ that are similar
distance or closer to RX J1230.8+0115 on the sky.

\section{Absorber Sizes as Inferred from Line Pair Statistics}

The only ``direct'' measurement of the physical sizes of \lya\ absorbers comes 
from the statistical analysis of line pairs in adjacent sight lines 
\citep[see e.g.,][]{Foltz:84,Dinshaw:97}.
In summarizing recent results, \citet{Dinshaw:98} showed that the statistics of 
line pairs
suggest characteristic {\it radii} for high column density (\lNHI $\geq 14.5$
\cm2; i.e., higher column density than the bulk of the \lya\ forest)
absorbers of 300-700\h70\ kpc for spherical clouds, with a possible trend 
towards larger sizes at smaller redshifts. The lowest redshift previously 
investigated using this technique is
one pair of sight lines at $z\sim$0.5 \citep{Dinshaw:97}. The maximum likelihood
analysis used on these pairs provides a range of radii extending more than a 
factor of two on either side of the values quoted above. However, all of the 
radius estimates are comparable to, or larger than, typical distances between 
bright galaxies. Thus, if these absorber radii are interpreted as being 
``contiguous'' structures, it is unlikely that \lya\ absorbers are physically or 
causally related to individual galaxies since their sizes are comparable to or 
larger than the distances between bright galaxies. These radii are more 
consistent with large-scale structure filaments found in numerical simulations
\citep{dave:99}. Additionally, the observation that not all absorbers found in 
one sight line have corresponding absorption in the other sight line implies that 
these are elongated structures rather than enormous spherical halos if the 
structures are ``contiguous.'' An alternative explanation is that coincident 
line pairs indicate ``correlated'' structures of smaller systems 
as in the case of galaxy halos aligned along a large-scale structure filament
\citep[see discussion in][]{Dinshaw:94,Dinshaw:97}. This is suggested in several
cases where substantial velocity and EW differences between coincident lines are
noted \citep{Dinshaw:94}.
 

\centerline{\epsfxsize=0.9\hsize{\epsfbox{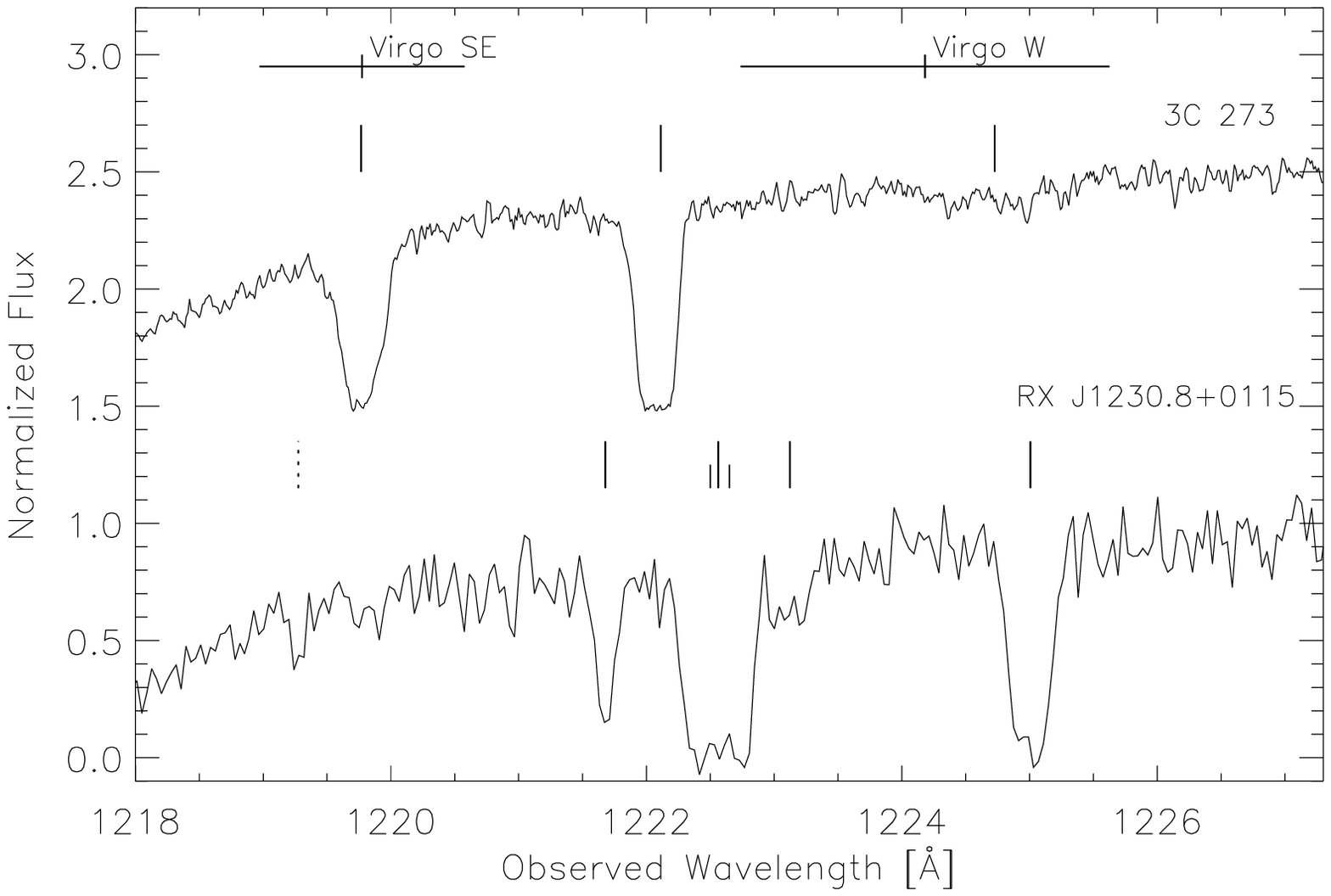}}}
\vspace{0.1in}
\figcaption{\label{fig:lya_lss} The 3C~273 (upper) and RX J1230.8+0115 (lower) 
spectra in the Virgo 
region. The ticks indicate the position of the absorbers (solid -- greater than
4-$\sigma$, dotted -- greater than 3.5-$\sigma$) in each sight line. The smaller 
ticks indicate the positions of the metal lines in the RX J1230.8+0115 spectrum. 
The crosses at the top of the
figure indicate the velocity ranges for the Virgo galaxy ``clouds," which
encompass most of the velocity space covered by the absorbers. There is also 
general correspondence in velocity between most of the absorption systems in 
the two lines-of-sight, yet the profiles and depths of the lines are extremely
different. \\}

The pairs analysis of these sight lines is consistent with the inferences 
obtained from other sight line pairs \citep{bechtold:94, crotts:94, 
Dinshaw:97, Dinshaw:98}. Figure 7 shows the 3C~273 (upper) and RX J1230.8+0115
(lower) spectra with the locations all $>4\sigma$ (solid ticks) and $>3.5\sigma$
(dotted tick) absorbers marked as well as 
the velocity range for the two Virgo Supercluster ``clouds'' of galaxies in 
this direction. We note that the
line pairs we have identified include absorbers with equivalent widths a factor 
of 4 less than those identified in the higher redshift studies of 
\citet{Dinshaw:97, Dinshaw:98}. Using the 
$\Delta v \leq$ 150 \kms\ criteria for ``coincident'' line pairs suggested by 
\citet{Dinshaw:97}, two of the three 3C~273 absorbers are coincident with
4-$\sigma$ absorbers in the other sight line and the third is coincident with a
marginal detection. For RX J1230.8+0115, four of five are coincident by the 
\citet{Dinshaw:97} definition by counting the 1685 and 1721 \kms\ absorbers as a 
single blended line and noting that the 1482 \kms\ absorber is also coincident 
with the same absorber in the 3C~273 sight line (thus, 3
``coincidences'' if this sort of double counting is not justified). In either case,
a large fraction of coincidences are present, suggesting that transverse sizes 
are large for at least some of these absorbers. 
The transverse distances between the sight lines at these redshifts 
is 200-500\h70\ kpc, assuming a
pure Hubble flow in this direction, comparable to those 
investigated using other sight line pairs. The lowest velocity absorbers beyond 
Virgo (and thus at higher redshifts than 
shown in Figure 7) have no coincident absorbers in the other sight line 
\citep{penton:00a} for transverse distances in the range 1.75-2.0\h70\ Mpc. These 
coincidences indicate characteristic \lya\ absorber {\it radii} of $\sim$ 100 -- 
250\h70\ kpc. Unlike at higher redshifts, no chance 
pairings are expected in this velocity interval, which could confuse these 
issues. Using the line densities found by
\citet{penton:00b} at low H~I column densities, we compute that there should be
0.14 $\pm$ 0.08 (a Poisson probability of 4$\times10^{-4}$) line pairs with
$\Delta v \leq$ 150 \kms\ created by chance in Figure 7.
Thus, all the absorber matches seen in
these two spectra are real (to 99.96\% confidence) statistical
associations of absorbers.

The only coincidence between the H~I and O~VI absorber at 1011 \kms\ in the 
3C~273 sightline and an RX J1230.8+0115 is with the marginal detection at
889 \kms. While this is only a single incidence, a size limit of $\leq$
225 h$^{-1}$ kpc (the distance between these two sightlines at this redshift)
supports the collisionally-ionized model for this absorber advocated
by T2002, because a photoionized model for this absorber requires a very large
physical size; i.e., we would expect a very large photoionized absorber to be
more uniform over these two sight lines. As shown by examples in Tripp et al. 
(2001) and Tripp and Savage 
(2000), the simultaneous presence of O~VI and absence of C~IV absorption may be 
a discriminator between collisionally-ionized and photoionized O~VI absorbers.

The characteristic radii derived above are based on using the $\Delta v \leq$ 150 
\kms\ criteria which was defined as a ``coincidence'' for much poorer resolution 
Faint Object Spectrograph (FOS) spectra ($\approx$ 25 times poorer than the 
present data). If these 
spectra were smoothed to the FOS resolution (Tripp, private communication), 
only one line pair match and two mismatches would be found -- the 
lower equivalent width lines in Figure 7 would not be detectable.
The remaining coincidence includes the three metal line absorbers at 1586 --
1721 \kms, suggesting a somewhat smaller but
still substantial, characteristic size for these structures. While the 
difference in these results is indicative of the resolution and sensitivity 
dependent nature of this measurement, we would still conclude that these
structures are large and elongated, comparable in size to the distance 
between these two sight lines.

Assuming that these absorbers reside in a gaseous filament with little turbulent 
motion, the Hubble flow velocity differences between absorbers in the two sight
lines should be in the range 30 -- 60 \kms. Therefore, despite the better STIS
resolution and sensitivity, the $\Delta$v $\leq 150$ \kms\
criterion for absorber coincidences still seems reasonable over such large
separations. While line pair matches within 150 \kms\ suggest a relationship 
between 
these absorber pairs, the velocities, equivalent widths, and profile shapes are 
not so well correlated as to require ``contiguous'' structures stretching across
these two sight lines (see e.g. Rauch et al. 2001 examination of sub-kpc
separated sight lines at high-$z$). However, we can not rule out the existence 
of a low column density (\lNHI $\sim 13$ cm$^{-2}$) contiguous structure, 
augmented at specific locations by higher column density gas; e.g., galaxy halos 
or winds embedded in a large-scale gaseous filament. The presence of a galaxy at 
similar velocity to the 1586 \kms\ absorber and at a small impact parameter from 
3C~273 (T2002 and Penton et al. 2002) may be evidence for this interpretation.

The small number of absorbers found in these two sight lines and the possible 
``special conditions'' of the Virgo Cluster  make statistical inference based 
upon this one sight line pair insecure. However, the approximate velocity 
coincidence between the metal line systems studied in detail here and in T2002 
strongly suggests that these absorbers are related. We discuss the 
sizes and shapes of the absorbers 
implied by these pairing further in the discussion section.

\section{Discussion and Conclusions}

The absorption line systems in the RX J1230.8+0115 and 3C~273 sight lines along 
the Southern edge of the Virgo Cluster provide a low redshift laboratory in which 
to study the properties of these absorbers and their relationship with their
environment. We have presented a detailed discussion of the physical properties
of these absorbers, concentrating on the characteristic sizes of low-$z$ \lya\ 
absorbers. We summarize these results in the next subsection. In \S 6.2, 
we present an interpretation of the nature of these absorbers inferred
from their properties and from the local environment of the \lya\ absorption
systems and present our conclusions about these absorbers.

\subsection {Properties of the Absorbers}

Based upon a detailed photoionization equilibrium calculation, the two blended 
absorbers in the RX J1230.8+0115 sight line at 1685 and 1721 \kms\ and the 1586 
\kms\ absorber in the 3C~273 sight line (T2002) have inferred sizes along the 
line-of-sight of $\sim$20 kpc and 70 pc respectively. These inferences are much 
more secure for the 3C~273 absorber because the hydrogen column density can be 
accurately determined due to available high quality FUSE
spectroscopy of the higher order Lyman lines. For the 3C~273 absorber,
T2002 find \lNHI = 15.85 \cm2, log($n_H$) = -2.8 $\pm$0.3 cm$^{-3}$, 
$Z$ = 0.06 $Z_\odot$ and an inferred thickness of $\sim$70 pc. For the case of 
the two blended absorbers studied here, the combination of HST STIS echelle and 
FUSE spectroscopy are insufficient to
provide similarly tight constraints. The FUSE spectrum is sufficient to rule out 
column densities as low as 10$^{15}$ \cm2\ and set a probable lower limit of 
\NHI = 10$^{16}$ \cm2\ for each absorber due to significant detections of 
Ly-$\gamma$ and Ly-$\delta$ in both components and  possible
detections of Ly-$\epsilon$, Ly-$\eta$ and Ly-$\theta$ in at least one component. 
Additionally, the low $b$-values that would be required
to fit \lya\ rule out column densities above \lNHI = 17.0 \cm2. 

We use the available STIS echelle and FUSE data to infer the physical conditions 
of these absorbers. The metal line ratios (N(C~IV)/N(C~II) and N(Si~IV)/N(Si~II)) 
constrain the ionization parameter (-2.85$<$log $U<$-2.60) for our best-guess
model, assuming the gas is photoionized. Assuming a standard value for the
extragalactic ionizing radiation intensity, we can translate this to limits on
the density of the gas. If we knew the 
metallicity of these structures, this would constrain the hydrogen column density 
and thus the absorber size along our line-of-sight. By 
assuming the same metallicity for these absorbers as for the 3C~273 
absorber located some 350\h70\ kpc away and column densities of \lNHI =
16 -- 17, consistent with constraints from the STIS and FUSE data, we
calculate line-of-sight sizes of 10 -- 30 kpc. Since these metallicities and 
column densities are similar to those obtained by T2002, the larger physical 
sizes obtained here are due to the higher ionization (lower density) of the 
RX J1230.8+0115 absorbers (C~IV is detected here, but not in 3C~273). The 
blended absorbers studied here have inferred line-of-sight sizes a factor of 
7 -- 300 times less than the minimum transverse size 
based upon detecting absorption in both sight lines at comparable redshifts,
while the 1586 \kms\ absorber in 3C~273 has a size that is a factor of
$\sim$4300 smaller than the minimum transverse size. This much larger inferred
transverse size is more suggestive of objects associated with large scale 
structure than with individual galaxies, while the smaller line-of-sight sizes 
are more consistent with structures related to individual galaxies (e.g., very 
extended halos or outflowing winds).
 
While only a few high column density absorbers similar to the ones in these 
two  sight lines are known, they are not unique at low-$z$. There is an 
absorber with a column density of \lNHI\ $\sim$16 \cm2\ at $z=0.167$ 
towards PKS 0405-1219 \citep{chen:00} and one with a column density of 
\lNHI\ = 15.57 at $z = 0.22497$ towards H1821+643 \citep{tripp:00,oegerle:00}.
Both of these systems very closely resemble the RX J1230.8+0115 $z = 0.005668$ 
complex in that they all contain a saturated absorber with a two component metal 
line system. These absorbers also have nearby galaxies, there is an
0.02 L$^*_K$ spiral galaxy at 90\h70\ kpc and a 1.2 L$^*_K$ elliptical at 
110\h70\ kpc from PKS 0405-1219 and there is a galaxy 112\h70\ kpc away from the 
H1821+643 absorber. The 
difference between the RX J1230.8+0115 $z = 0.005668$ complex and these systems 
is the existence of higher ionization species detected with O~VI and N~V 
detected in PKS 0405-1219 and O~VI and Si~III detected in H1821+643. 

At somewhat higher redshifts, \citet*{rigby:02} identify a class of weak Mg~II 
systems which have similar physical properties to the absorbers discussed
above including sizes on the  order of 10
pc to several kpc. For the limited number of regions around the
absorbers that have been surveyed optically, there are no galaxies down
to 0.01L$^*$ within 70\h70\ kpc, but larger regions around the
absorbers have not been studied. 

While the lower ionization weak Mg~II absorbers are most similar to the absorbers 
studied here and in the 3C~273 sight line, there are also more highly ionized metal 
line absorbers (i.e., O~VI detected) that are found somewhat farther 
from galaxies, but that seem to be associated with galaxy groups. The PKS 
2155-304 sight line has 6 absorbers between 16,000 and 
18,000 \kms, which appear to be associated with a small group of spiral galaxies, 
although none of these are within 400\h70\ kpc of the absorbers \citep{shull:98}. 
Two of these absorbers have been detected in O~VI with FUSE \citep{Shull:03} 
and there is a tentative detection of O~VIII with CHANDRA \citep{Fang:02}. The
two O~VI absorbers in the PG 0953+415 sight line \citep{Savage:02} fall near 
peaks in the galaxy density, with the $z = 0.14232$ absorber having 4 galaxies
within 130 \kms\ and 3 Mpc, the closest at a projected distance of 395 kpc
\citep{trippsavage:00}. Similarly, the H1821+643 $z = 0.1212$ absorber appears to 
fall within a galaxy group, with 7 galaxies within 500 \kms\ and 3.1 Mpc. These
higher ionization absorbers have larger inferred line-of-sight sizes than those 
discussed above (based on a photoionization model). However, since there are no 
adjacent sight lines, there are no constraints on their transverse sizes.

We note that the column densities of the absorbers studied here 
are unusually high for the bulk of the local \lya\ forest absorbers and so are 
probably not representative of the properties (size, metallicity, etc.) or the 
environment of the lower column density lines (\lNHI $< 14$ \cm2; Penton et al.
2002). 
  
\subsection{Galaxy/Absorber Associations and Large Scale Structure}

\citet{penton:02} examined the hypothesis that the 3C~273 and RX J1230.8+0115 
absorbers at Virgo distance are due to extended galaxy halos and concluded that 
the evidence was inconsistent with the available data. Based upon the known 
galaxy redshifts, the nearest galaxies to the strong absorbers in these sight 
lines were $\geq$200\h70\ kpc away and there seemed to be no correlation between 
impact parameter and the probability of detecting absorption as would be
expected in a simple halo model. The galaxy redshift surveys have been improved
and a dwarf galaxy has been found only 71\h70\ kpc from the 3C~273 1586 \kms\ 
absorber \citep{stocke:03}. Finding this galaxy may indicate that the halos
of dwarf galaxies or outflows from them are responsible for some \lya\
absorbers. The ``post-starburst'' signature in the optical spectrum of this 
galaxy makes an
outflow model for the 3C~273 absorber particularly attractive since it can 
explain the sub-solar metal abundances as well as the super-solar (Si/C) ratio,
which is indicative of SN type II enrichment. 
 
The RX J1230.8+0115 $z = 0.005668$ complex could be associated with a gas-rich
galaxy (H I 1225+01) 400 \kms and 270\h70\ kpc away, or with the edge-on dwarf 
spiral (CGCG 014-064) 280\h70\ kpc and only 56 \kms\ away. 
The fact that the dwarf galaxy 71\h70\ kpc from 3C~273 was missed by 
\citet{morris:93} provides a warning that the RX J1230.8+0115 region may not have 
been surveyed sufficiently well  to conclude that there are no galaxies of 
similar luminosity near these blended absorbers, although there have been a
number of relatively deep, independent surveys of this region. If another 
faint galaxy is found close to RX J1230.8+0115 at $cz\sim$1700 \kms, then the 
model of outflowing winds from dwarf galaxies \citep{stocke:03} is an exceptional 
fit for both the 3C~273 and RX J1230.8+0115 $cz\sim$1700 \kms\ absorbers 
\citep{stocke:03}. The higher than average galaxy density ($\approx 50$ times
the average) in this region at the 
southern edge of the Virgo Cluster make this a reasonable possibility. However, 
this scenario may require very large amounts of H I associated with quite 
small galaxies (see below and Stocke et al. 2003 for further discussion).

RX J1230.8+0115 is 0.9$\deg$ from 3C~273, and both of these lines-of-sight
pass through a filament of galaxies identified by \citet{penton:02} as well as
passing through the Southern Extension of Virgo and near the W cloud \citep[the 
W' cloud, discussed as part of this region by T2002 lies several degrees 
to the north;][]{binggeli:93}. These galaxy clouds are thought to be 
independent of the Virgo Cluster. The W cloud is a large feature that may be
another filament stretching to the north at higher velocities than are found in
the \citet{penton:02} filament. However, the SE cloud is embedded in the 
northern end of this filament. \citet{penton:02} showed that both of the strong 
3C~273 Virgo absorbers, all of the RX J1230.8+0115 Virgo absorbers, plus the 
seven (now eight) known galaxies in the region all lie along a single large-scale 
structure filament that could extend for many Mpc crossing these sight lines.  

The filamentary model for these absorbers is consistent with the abundance of 
common velocity ($\pm$ 150 \kms) absorber pairs seen in these two sight lines 
but {\it not} with the line-of-sight thicknesses of these structures inferred 
from photoionization models. Either the methods used to estimate 
absorber sizes could be wrong, a very thin ``ribbon'' or ``sheet'' of gas 
could extend along this filament of galaxies for at least several hundred kpc, 
or, because the pairs analysis cannot distinguish
between ``contiguous'' or ``correlated'' structures as emphasized by its
practitioners \citep{Dinshaw:97}, these absorbers map out correlated rather 
than contiguous structures. Given the difference in depth versus transverse 
size for these absorbers, and the lack of detailed correlations between the 
absorber properties (i.e., velocity differences greater than $\pm$ 50 \kms, 
which could be accounted for by Hubble flow, as well as large equivalent width 
differences; see \S 5) in these two sight lines, a correlated 
structure model seems much more plausible. One possibility is that the absorbers 
in this correlated structure are related to individual galaxies whose halos or 
outflowing winds have caused significant clumping and metal enrichment in small 
regions of the large gaseous structure in which they are
embedded. If this is the case, we might expect to find 
one or more faint galaxies close to the RX J1230.8+0115 Virgo absorbers. 

The high resolution STIS echelle spectra, shown in Figure 7, indicate that 
there is no detailed correlation between the 
velocities or the equivalent widths of the absorbers in these sight lines. 
These differences in velocity and equivalent width structure contrast with
the extremely detailed velocity and equivalent width correlations between the 
Ly$\alpha$ absorbers in the two  gravitationally-lensed images of Q1422+231 
\citep{Rauch:01} separated by 80h$^{-1}_{70}$ parsecs at $z = 3.3$. In the 
case of the Q1422+231 absorbers there is a detailed agreement between these 
absorber properties, requiring a quiescent, contiguous gaseous structure 
\citep{Rauch:01}. The differences between the
Q1422+231 sight lines and these sight lines may indicate the
difference in scale over which we see contiguous structures (tens of parsecs) 
and correlated structures (hundreds of kpc).

If we assume that the absorber depths of 1 -- 30 kpc calculated in \S 3
and in T2002 are indicative of the overall three-dimensional size of these
absorbers, then, in order to obtain a covering factor of unity (as
would be required to detect three such absorbers in the two sight lines
observed), there would need to be 10 to 10,000 of these structures in the
filament (assuming a filament thickness of $\sim$ 100 kpc). 
Charlton et al. (2002) also find that a large number of weak Mg~II absorbing
structures are required to explain their detected population found at higher 
redshifts; i.e., these absorbers must greatly outnumber bright galaxies. 
Although not yet observed at Mg~II, the three absorbers discussed here are 
very similar to those studied by \citet{Charlton:02}. If these absorbers are 
associated with galaxies, we would expect to find 
numerous faint or low surface brightness galaxies throughout this filament. 
While we make the case in \S 6 that some galaxies may have been missed, 
this filament has been thoroughly surveyed for galaxies and only 8 have been 
found. It is very hard to believe that large numbers of star formation
sites have been missed given the volume of galaxy redshift data available for
this region (\S 4).

One possible scenario is that while these absorbers have small thicknesses, they 
could be quite extended on the sky if the metal enriched gas is in the form of a 
thin shell surrounding a galaxy, perhaps driven by a wind. In this case, the area 
on the sky 
would be substantially larger than the thickness derived from photoionization; 
e.g., the radius of the shell could extend 50 -- 100h$^{-1}$ kpc from the galaxy 
if the distance to the galaxy near the 3C~273 absorber is indicative of the size 
of these shells (impact parameter $= 71 h^{-1}$ kpc; \S 4 and Stocke et al. 2003). 
In this case, only a few ($\sim$4) gaseous shells would be required to have the 
covering fraction of unity necessitated by the 3C~273/RX~J1230.8+0115
observations. In the shell scenario, it 
would also be quite natural to find two absorbers closely spaced in velocity along 
a single sight line (e.g., RX~J1230.8+0115) due to the interception of both sides
of the expanding shell of gas. For a 75 kpc radius shell with an expansion 
velocity of 18 \kms\ (the separation between the RX~J1230.8+0115 metal lines) the 
epoch of star formation would have occured $\sim$ $4\times10^9$ yrs
ago expelling $7\times10^9$ \msolar\ of gas (given the average RX~J1230.8+0115
gas density and shell thickness). Thus, the galaxy putatively associated with 
the enriched gas in these absorbers does not need to be currently forming 
stars. This is the case for the dwarf near the 3C~273 absorber
\citep{stocke:03}, which would have had to expel $\sim2\times10^8$ \msolar\ of
gas in this scenario. This scenario requires large masses of gas to have been
expelled from galaxies, but may be compelling if another galaxy is found near 
RX~J1230.8+0115. Given the significant amount of optical spectroscopy already 
expended on this region, it may be more likely that the 
column density of the absorber is on the high end of our range, 
making the metallicity significantly smaller. In this lower metallicity
scenario, these systems could represent more 
primordial overdensities expected to be associated with large scale structure
\citep{dave:99,stocke:01,penton:02}. Given the proximity of these absorbers to 
us, this region is the best available laboratory with which to test this 
hypothesis so we are presently surveying the region at the 21cm line of H~I to 
look for gas-rich galaxies as well as continuing our optical spectroscopy to 
answer this question definitively.
   
\acknowledgments

JLR thanks Steve Penton for assistance extracting and calibrating the RX
J1230.8+0115 STIS spectrum and Ken Sembach for input on the FUSE spectrum. The
authors thank Todd Tripp for an extremely careful reading of the paper and
helpful suggestions. JTS, MLG and JLR acknowledge financial support from 
NASA-HST Archival Research Grant AR-09221.01-A for this work. JTS also 
acknowledges support for the Cosmic Origins Spectrograph GTO Team through NAG 
512279. JTS thanks Ray Weymann and Kevin McLin for permission to quote some 
results from our galaxy survey work in the RX~J1230.8+0115 and 3C~273 region 
prior to publication.

\clearpage
\begin{deluxetable}{crcccrccrc}
\tablewidth{6.8in}
\small
\tablecaption{Measurements of Virgo Redshift Lyman-$\alpha$ Absorbers in RX
J1230.8+0115}
\tablehead{
\colhead{$\lambda$} & \colhead{Vel.} &\colhead{$z$} & \colhead{$\Delta z$} &
\colhead{b} & \colhead{$\Delta$ b} & \colhead{$\log$ N$_{HI}$} &
\colhead{$\Delta \log$ N$_{HI}$} & \colhead{EW} & \colhead{4$\sigma$ EW}\\
\colhead{(\AA)} & \colhead{(\kms)} & \colhead{} & \colhead{} & \colhead{(\kms)} &
\colhead{(\kms)} & \colhead{(cm$^{-2}$)} & \colhead{(cm$^{-2}$)}
& \colhead{(m\AA)} & \colhead{(m\AA)}}
\startdata
1219.27 &  889 & 0.002964 & 0.000012 & 17  &  5 & 13.10 &  0.12  &  46 &  52 \\
1221.68 & 1482 & 0.004945 & 0.000004 & 22  &  2 & 13.68 &  0.04 & 158 &  56 \\
1222.50\tablenotemark{1} & 1685 & 0.005621 & \nodata  & 23  &  1 & 16.20 &  \nodata &
497\tablenotemark{2} & 54 \\
1222.65\tablenotemark{1} & 1721 & 0.005742 & \nodata  & 17  &  1 & 16.60 &  \nodata &
410\tablenotemark{2} & 44 \\
1223.11 & 1834 & 0.006120 & 0.000023 & 53  & 11 & 13.47 &  0.09 & 115 & 56 \\
1225.01 & 2302 & 0.007680 & 0.000004 & 33  &  2 & 14.30 &  0.06 & 360 & 69 \\
\enddata

\tablenotetext{1}{These values represent the ``best-fit" model of the
$z = $0.005668 saturated blend, constrained only by the positions of the metal
lines (see \S 2.2).}
\tablenotetext{2}{These EW values are derived from the Voigt profile fit to the
line, since they can not be separated non-parametrically in these data.}

\end{deluxetable}

\begin{deluxetable}{lccccrcccrr}
\tablewidth{7.2in}
\small
\tablecaption{Measurements of Metal Lines in the $z = 0.005668$ Absorber}
\tablehead{
\colhead{Ion} & \colhead{} &
\colhead{Vel.} &\colhead{$z$} & \colhead{$\Delta z$} & \colhead{b} &
\colhead{$\Delta$ b} & \colhead{$\log$ N$_{ION}$} &
\colhead{$\log \Delta$N$_{ION}$} & \colhead{EW} & \colhead{4$\sigma$ EW}\\
\colhead{} & \colhead{} & \colhead{(\kms)} &\colhead{} & \colhead{} &
\colhead{(\kms)} & \colhead{(\kms)} & \colhead{(cm$^{-2}$)} &
\colhead{(cm$^{-2}$)} & \colhead{(m\AA)} & \colhead{(m\AA)}}
\startdata
C II    &  & 1685 & 0.005621 &  0.000003 &  6 &  2 & 13.22 & 0.07 & 29 & 21 \\
C II    &  & 1721 & 0.005742 &  0.000002 & 11 &  1 & 13.69 & 0.03 & 71 & 23 \\
C IV    &  & 1679 & 0.005600 &  0.000012 & 15 &  6 & 12.88 & 0.17 &
24\tablenotemark{1} & 30 \\
C IV    &  & 1720 & 0.005738 &  0.000003 & 10 &  2 & 13.26 & 0.06 &
66\tablenotemark{1} & 33 \\
Si II   &  & 1687 & 0.005628 &  0.000014 & 19 &  7 & 12.42 & 0.14 & 27 & 26 \\
Si II   &  & 1720 & 0.005738 &  0.000004 &  7 &  2 & 12.49 & 0.09 & 40 & 31 \\
Si IV   &  & 1684 & 0.005617 &  0.000015 & 14 &  7 & 12.46 & 0.20 &
16\tablenotemark{2} & 25 \\
Si IV   &  & 1718 & 0.005731 &  0.000004 &  8 &  2 & 12.86 & 0.08 &
43\tablenotemark{2} & 26 \\
\enddata

\tablenotetext{1}{This is the EW for the C~IV 1548 line}
\tablenotetext{2}{This is the EW for the Si~IV 1393 line}

\end{deluxetable}

\begin{deluxetable}{llrrrc}
\tablewidth{5.3in}
\tablecaption{Distances from Galaxies in the RX J1230.8+0115 and 3C 273 Fields}
\tablehead{
\colhead{Absorption} & \colhead{$z_{abs}$} & \colhead{V$_{abs}$} & \colhead{$\Delta$ V} &
\colhead{D$_{perp}$} & \colhead{Nearest} \\
\colhead{System} & \colhead{} & \colhead{[km s$^{-1}$} & \colhead{[km s$^{-1}$]} &
\colhead{[Mpc]} & {Galaxy}}
\startdata

RX J1230.8+0115-1  & 0.002963 &  888  & -218 & 0.117 & CGCG 014-054 \\
RX J1230.8+0115-2  & 0.003437 &  1030 &  -76 & 0.117 & CGCG 014-054 \\
RX J1230.8+0115-3  & 0.004945 &  1482 &  184 & 0.271 & HI 1225+01 \\
RX J1230.8+0115-4  & 0.005668 &  1699 &   56 & 0.279 & CGCG 014-064 \\
RX J1230.8+0115-5  & 0.006148 &  1843 &  200 & 0.279 & CGCG 014-064 \\
RX J1230.8+0115-6  & 0.007680 &  2302 &   13 & 0.162 & [ISI96]1228+0116 \\
\cline{1-6}
3C 273-1           & 0.00337  &  1010 &  -96 & 0.104 & CGCG 014-054 \\
3C 273-2           & 0.00530  &  1589 &   33 & 0.071 & New galaxy \\
3C 273-3           & 0.00745  &  2233 &  -96 & 0.396 & Leda 139866 \\

\enddata
\end{deluxetable}

\end{document}